\documentclass[sigconf]{acmart}

\usepackage[T1]{fontenc}

\usepackage{multirow}
\usepackage{amsmath}
\usepackage{listings}
\usepackage{enumitem}

\colorlet{punct}{red!60!black}
\definecolor{background}{HTML}{EEEEEE}
\definecolor{delim}{RGB}{20,105,176}
\colorlet{numb}{magenta!60!black}

\lstdefinelanguage{json}{
    basicstyle=\normalfont\ttfamily,
    numberstyle=\scriptsize,
    breaklines=true,
    stepnumber=1,
    numbersep=8pt,
    showstringspaces=false,
    frame=lines,
    backgroundcolor=\color{background},
    literate=
     *{0}{{{\color{numb}0}}}{1}
      {1}{{{\color{numb}1}}}{1}
      {2}{{{\color{numb}2}}}{1}
      {3}{{{\color{numb}3}}}{1}
      {4}{{{\color{numb}4}}}{1}
      {5}{{{\color{numb}5}}}{1}
      {6}{{{\color{numb}6}}}{1}
      {7}{{{\color{numb}7}}}{1}
      {8}{{{\color{numb}8}}}{1}
      {9}{{{\color{numb}9}}}{1}
      {:}{{{\color{punct}{:}}}}{1}
      {,}{{{\color{punct}{,}}}}{1}
      {\{}{{{\color{delim}{\{}}}}{1}
      {\}}{{{\color{delim}{\}}}}}{1}
      {[}{{{\color{delim}{[}}}}{1}
      {]}{{{\color{delim}{]}}}}{1},
}

\DeclareMathOperator*{\argmax}{argmax} 

\newcommand{\miniskip}{\vspace*{-.5\baselineskip}}
\newcommand{\shrink}{\vspace*{-.9\baselineskip}}

\copyrightyear{2020}
\acmYear{2020}
\setcopyright{acmcopyright}
\acmConference[SIGIR '20]{Proceedings of the 43rd International ACM SIGIR Conference on Research and Development in Information Retrieval}{July 25--30, 2020}{Virtual Event, China}
\acmPrice{15.00}
\acmDOI{10.1145/3397271.3401416}
\acmISBN{978-1-4503-8016-4/20/07}
\begin{document}

\title{REL: An Entity Linker Standing on the Shoulders of Giants} 

\author{Johannes M. van Hulst}
\affiliation{Radboud University}
\email{mick.vanhulst@gmail.com}

\author{Faegheh Hasibi}
\affiliation{Radboud University}
\email{f.hasibi@cs.ru.nl}

\author{Koen Dercksen}
\affiliation{Radboud University}
\email{koen.dercksen@ru.nl}

\author{Krisztian Balog}
\affiliation{University of Stavanger}
\email{krisztian.balog@uis.no}

\author{Arjen P. de Vries}
\affiliation{Radboud University}
\email{a.devries@cs.ru.nl}

\renewcommand{\shortauthors}{J.M. van Hulst, et al.}

\begin{abstract}
Entity linking is a standard component in modern retrieval system that is often performed by third-party toolkits.  Despite the plethora of open source options, it is difficult to find a single system that has a modular architecture where certain components may be replaced, does not depend on external sources, can easily be updated to newer Wikipedia versions, and, most important of all, has state-of-the-art performance.
The REL system presented in this paper aims to fill that gap.
Building on state-of-the-art neural components from natural language processing research, it is provided as a Python package as well as a web API.  We also report on an experimental comparison against both well-established systems and the current state-of-the-art on standard entity linking benchmarks.
\end{abstract}

% The code below should be generated by the tool at http://dl.acm.org/ccs.cfm
%\begin{CCSXML}
%<ccs2012>
%<concept>
%<concept_id>10002951.10003317.10003347.10003352</concept_id>
%<concept_desc>Information systems~Information extraction</concept_desc>
%<concept_significance>500</concept_significance>
%</concept>
%</ccs2012>
%\end{CCSXML}
%
%\ccsdesc[500]{Information systems~Information extraction}


\maketitle

\miniskip
\section{Introduction}
\label{sec:intro}

%\paragraph{Why entity linking is important?}
%\begin{itemize}
%	\item Improvements obtained by entity linking for doc retrieval and entity retrieval
%	\item EL for query understanding in Web and conversational search
%	\item In general: EL is used for downstream processing for many IR taskshttps://www.sharelatex.com/learn/Positioning_of_Figures
%	\item An effective and efficient entity linker is needed by both academia and industry (I am thinking of companies like Signal)
%\end{itemize}

Entity linking (EL) refers to the task of recognizing mentions of specific entities in text and assigning unique identifiers to them from an underlying knowledge repository~\citep{Balog:2018:EOS}.  The problems of entity recognition and disambiguation have traditionally been studied in the natural language processing (NLP) community.  It was also them who first recognized the utility of Wikipedia as a large-scale knowledge repository to disambiguate against~\citep{Bunescu:2006:UEK,Cucerzan:2007:LSN}.
This line of work has been quickly followed up by information retrieval (IR) researchers~\citep{Milne:2008:LLW,Mihalcea:2007:WLD}.
Over the past years, entity linking has become a standard component in modern retrieval systems, and has been leveraged in a range of tasks, including document ranking~\citep{Xiong:2017:WDR}, entity retrieval~\citep{Hasibi:2016:EEL}, knowledge base population~\citep{Balog:2013:MCA}, and query recommendation~\citep{Reinanda:2015:MRR}.  % also query interpretation~\citep{Carmel:2014:EER,Hasibi:2015:ELQ} => but the focus there is on EL, so better leave that out
Since entity linking is not the main focus of these works, it is commonly performed by some third-party toolkit, with the resulting annotations being utilized in downstream processing.  
Some of the most prominent toolkits used for this purpose include DBpedia Spotlight~\citep{Mendes:2011:DSS}, TAGME~\citep{Ferragina:2010:TOA}, WAT~\cite{Piccinno:2014:WAT}, and FEL~\cite{Pappu:2017:LME}.

%\paragraph{What are the problems with existing tools?}
%\begin{itemize}
%	\item All freely available/open source entity linkers (DBpedia Spotlight, TAGME, WAT, FEL) are behind the state-of-the-art entity linkers, not incorporating recent NLP progress (e.g., embeddings) 
%	\item Yahoo Fast entity linker (FEL) is not maintained any more
%	\item SMAPH is using external source (search engines)
%	\item Nordlys: meant for short text, not efficient for long text
%\end{itemize}

Existing toolkits fall short in a number of areas.  Some are unmaintained~\cite{Pappu:2017:LME}; others are meant for short text and inefficient for long text~\cite{Hasibi:2017:NTE}; some rely on external sources like web search engines~\cite{Cornolti:2018:APA}. Typically, they are shipped with a specific Wikipedia version that has become dated, causing difficulties when attempting to update to a recent Wikipedia~\cite{Piccinno:2014:WAT, Cornolti:2018:APA}. An issue that is often not addressed is the lack of speed (throughput). Most importantly, none of the default open source entity linkers incorporate recent progress made in the NLP community on neural network-based approaches~\cite{Kolitsas:2018:ENE}. With this work, we aim to close that gap and remedy all of these problems by introducing an efficient, up-to-date entity linker that has a modular architecture to ease, e.g., updates of external resources like Wikipedia.

%\paragraph{Why REL ?}
%

We present REL\footnote{REL in Dutch means mayhem, interference, or
  disturbance; and, it is easily recognized to abbreviates `relatie' (relation in English).}
 (which stands for Radboud Entity Linker), an open source toolkit for entity linking. 
REL stands on the shoulders of giants and is an ensemble of multiple methods and packages from the state-of-the-art natural language processing research. REL has been developed with the following design considerations:
%\minishrink
%
\begin{itemize}[leftmargin=3mm]
	\item Use \emph{state-of-the-art} approaches for entity disambiguation (ED)~\cite{Le:2018:IEL, Ganea:2017:DJE} and named entity recognition (NER)~\cite{Akbik:2018:flair}, ensuring it is on par with the state-of-the-art on end-to-end entity linking~\citep{Kolitsas:2018:ENE}.
	\item Use a \emph{modular architecture} with mention detection (using a NER approach) and entity disambiguation components. Specifically, separating mention detection from entity disambiguation enables us to choose an NER method appropriate for the context in which entity linking is employed (i.e., optimizing for recall vs. throughput). % e.g., high recall vs. improved efficiency).  %(e.g., an n-gram-based approach for high recall or spaCy~\todo{[REF]} for improved efficiency).
	\item Design for sufficient \emph{throughput}; reporting 700 ms for an average document of 300 words.  Notably, most of the time is used for NER, which could be changed to a more efficient option.%, e.g., SpaCy, by sacrificing recall.%), while disambiguation is only around \todo{0.06 sec}. 
	\item Develop a \emph{lightweight} solution that can be deployed on an average laptop/desktop machine; it does not need much RAM, and, importantly, it does not need a GPU.  %: due to use of SQLite, it does not need high RAM, also no need for GPU. The system can be simply deployed on a laptop/local machine.
%	\item \mi{NER step can be optimised by utilising a GPU (see table efficiency) and in the near future we plan to implement a more efficient batch loading method. For larger corpuses, such as Washington post, we can preload a set of documents and parse them at the same time. Example: parsing a batch 20 sentences (i.e. single document) takes the NER-tagger (on GPU) .08 sec, parsing a concatenated set of 10 documents, totaling 191 sentences takes .42 seconds.}
	\item Train on a recent Wikipedia dump (2019-07) and ensure easy \emph{updates} to new Wikipedia versions (all necessary scripts included).  %; few commands are needed to train the embeddings and a feedforward neural models (for disambiguation)
	%\item It can be deployed as a Python package or used via a restful API.
\end{itemize}
REL is available at \url{http://tiny.cc/RadboudEL} under a MIT license, can be deployed as a Python package, or used via a restful API.

% PS. REL stands for Radboud Entity Linker. REL in Dutch means riot, but more in the sense to 'stir things up'. Could be suitable, because tagme has been the standard tool for a long time and we aim to change that ... (suggestions are welcome :))

\section{Entity linking in REL} 
\label{sec:approach}
In this section, we present the entity linking method underlying REL.  We follow a standard entity linking pipeline architecture~\citep{Balog:2018:EOS}, consisting of three components: (i) mention detection, (ii) candidate selection, and (iii) entity disambiguation.

\miniskip
\subsection{Mention Detection}
\label{sec:approach:md} 
In the mention detection step, we aim to detect all text spans that can be linked to entities. These text spans, referred to as \emph{mentions}, are obtained by employing a Named Entity Recognition (NER) tool. NER taggers detect entity mentions in text and annotate them with (coarse-grained) entity types~\cite{Balog:2018:EOS}.  We employ Flair~\cite{Akbik:2018:flair}, a state-of-the-art NER based on contextualized word embeddings.  Flair takes the input to be a sequence of characters and passes it to a bidirectional character-level neural language model to generate a contextual string embedding for each word.  These embeddings are then utilized in a sequence labeling module to generate tags for NER.
 
Using a NER method for mention detection enables us to strike a balance between precision and recall. Another approach, which may result in high recall,
% \todo{$\Leftarrow$ I’m not sure the dictionary-based approach would provide high recall. I’d perhaps be more safe to say that it can be more efficient than NER-based mention detection.} 
 is matching all n-grams (up to a certain $n$) in the input text against a rich dictionary of entity names~\cite{Hasibi:2015:ELQ,Balog:2018:EOS}.  In REL, the mention detection component can easily be replaced by another NER tagger such as spaCy\footnote{\url{https://spacy.io/}} or by a dictionary-based approach. %, given that the output adheres to the format defined by REL.
 
 \miniskip
 \subsection{Candidate Selection}
 \label{sec:approach:cs} 
 For each text span detected as a mention, we select up to $k_1+k_2$ (=7) candidate entities (following~\cite{Ganea:2017:DJE}). The $k_1$ (=4) candidate entities are selected from the top ranked entities based on the mention-entity prior $p(e|m)$, for a given entity $e$ and a mention $m$. To compute this prior, we sum up hyperlink counts from Wikipedia and from the CrossWikis corpus~\cite{Spitkovsky:2012:CDE} to estimate probability $P_{\text{Wiki}}(e|m)$. A uniform probability $P_{\text{YAGO}}(e|m)$ is also extracted from YAGO dictionary~\cite{Hoffart:2011:RDN}. These two probabilities are combined into the final $P(e|m)$ prior as $\mathrm{min}(1, P_{\text{Wiki}}(e|m) + P_{\text{YAGO}}(e|m))$~\cite{Ganea:2017:DJE}.
 
 The other $k_2$ (=3) candidate entities are chosen based on their similarity to the context of the mention. This similarity score is obtained by $\textbf{e}^T \sum_{w \in c} \textbf{w}$, where $c$ is n-word ($n=50$) context surrounding mention $m$ and $\textbf{w}$ and $\textbf{e}$ are entity and word embedding vectors. This score is computed for $k$ (=30) entities with the highest $P(e|m)$ prior and the top-$k_2$ entities are added to the list of candidate entities~\cite{Ganea:2017:DJE}.

%\mi{Candidate selection is actually reversed. We first select three candidates based on embedding scores, after which we choose the remaining four based on $P(e|m)$.}
 
In REL, we use Wikipedia2Vec word and entity embeddings~\cite{Yamada:2016:JLE} to estimate the similarity between an entity and a mention's local context. Wikipedia2Vec jointly learns word and entity embeddings from Wikipedia text and link structure, and is available as an open source library~\cite{Yamada:2018:WOT}. The hyper-parameters $k_1$, $k_2$, $k$, and $n$ are set based on the recommended values in~\cite{Le:2018:IEL, Ganea:2017:DJE}.
%\todo{(Consider replacing 50 and 30 with variables, like with $k_1$ and $k_2$. You may also want to add a sentence on why/how these specific values are chosen; based on preliminary experiments, prior work, etc. I'd use the variables in the previous paragraph, and the define the specific values here, just like you did with the choice of embedding used.)}

\miniskip
 \subsection{Entity Disambiguation}
 \label{sec:approach:ed} 
In the entity disambiguation step, we link mentions to their corresponding entities in the knowledge graph (here: Wikipedia). Entity disambiguation in REL is based on the Ment-norm method proposed by \citet{Le:2018:IEL}. Given an input document $D$, the entity linking decisions are made by %combining the local and global coherence scores:
combining local compatibility (which includes prior importance and contextual similarity) and coherence with the other entity linking decisions in the document:
\begin{equation}
	E^* = \argmax\limits_{E \in C_1 \times ... \times C_n} \sum\limits_{i=1}^{n} \psi(e_i, c_i) + \sum\limits_{i \neq j} \phi(e_i, e_j, D) ~,
	\label{eq:local_golobal}
\end{equation}
where $C_i$ denotes the set of candidate entities for mention $m_i$ and $E=\{e_1, .., e_n\}$. The coherence score between entity $e_i$ and its local context $c_i$ is computed by the function $\psi(e_i, c_i)$ as defined in~\citep{Ganea:2017:DJE}, and the coherence between all entity linking decisions is captured by the function $\phi(e_i, e_j, D)$. \citet{Le:2018:IEL} compute the $\phi$ function by incorporating relations between mentions of a document. Assuming $K$ latent relations, $\phi$ is calculated as:
\begin{equation}
	\phi(e_i, e_j, D) = \sum\limits_{k=1}^K \alpha_{ijk} \textbf{e}_i^T \textbf{R}_k \textbf{e}_j ~,
\end{equation}
where $\mathbf{e}_i, \mathbf{e}_j \in \mathbb{R}^d$ are the embeddings of entities $e_i, e_j$ (using the same embeddings as in the candidate selection step), $\mathbf{R}_k$ is a diagonal matrix, and $\alpha_{ijk}$ is a normalized score defined as: 
\begin{equation}
	\alpha_{ijk} = \frac{1}{Z_{ijk}} \exp\Big\{\frac{f^T(m_i, c_i) \textbf{D}_k f(m_j, c_j)}{\sqrt{d}}\Big\} ~,
\end{equation}
where $\textbf{D}_k \in \mathbb{R}^{d \times d}$ is a diagonal matrix, and function $f$ is a single-layer neural network that maps mention $m_i$ and its context $c_i$ to a $d$-dimensional vector. $Z_{ijk}$ is a normalization factor over $j$ and is computed as:
\begin{equation}
Z_{ijk} = \sum\limits_{\substack{j'=1 \\ j' \neq i}}^n \exp\Big\{\frac{f^T(m_i, c_i) \textbf{D}_k f(m_j, c_j)}{\sqrt{d}}\Big\} ~.
\end{equation}
The optimization of Eq.~\eqref{eq:local_golobal} is performed using max-product loopy belief propagation (LBP), and the final score for an entity of a mention is obtained by a two-layer neural network that combines $P(e|m)$ with max-marginal probability of an entity for a given document.
% using a two-layer neural network. 
The training of the model, referred to as the \emph{ED model} henceforth, is performed using max-margin loss.
To estimate posterior probabilities of the linked entities, we fit a logistic function over the final scores obtained by the neural model~\cite{Platt:2000:PSVM}.

\section{Implementation and Usage}
\label{sec:Imp}
\vspace*{-0.25\baselineskip}
Next, we describe the implementation details and usage of REL.

\vspace*{-0.25\baselineskip}
\subsection{Implementation Details}

\noindent
\emph{\textbf{Memory and GPU usage.}} One of the design requirements of REL is being lightweight, such that it can be deployed on an average machine. To minimize memory requirements, we store Wikipedia2Vec entity and word embeddings, GloVe embeddings, and an index of pre-computed $P(e|m)$ values (i.e., a surface form dictionary) in a SQLite3\footnote{\url{https://www.sqlite.org/index.html}} database. Using SQLite, we are able to minimize memory usage for our API to 1.8GB if the user chooses to not preload embeddings. REL also does not require GPU during inference. The neural model used for entity disambiguation is a feed-forward network and does not require heavy CPU/GPU usage. Training of Wikipedia2Vec embeddings, however, requires high memory and is done more efficiently using a GPU.
%\todo{(Do you have some numbers here? How long does it take to train using what GPU configuration?)} \fa{Not really}
\\

%\todo{This part is a bit repetitive w.r.t. Sect. 2. The different sources used for computing $P(e|m)$ have already been described there; those may be removed from here.  The types of embeddings used should also probably be under methods.  Focus on design decisions behind the implementation and key details (like the $P(e|m)$ index). What does the ``$P(e|m)$ index'' exactly entail? It's a bit confusing. Do you mean pre-computed probabilities that are looked up in a database? You could also consider simply listing all the data sources used (with specific versions) in a separate subsection 3.1 Data Sources.}

\noindent
\emph{\textbf{REL components.}} REL has a modular architecture, with separate components for mention detection, entity disambiguation, and the generation of the $P(e|m)$ index. The mention detection component is based on the Flair package\footnote{\url{https://github.com/flairNLP/flair}} and can be easily replaced by another mention detection approach. The disambiguation component is implemented using PyTorch and based on the source code of~\citep{Le:2018:IEL}.\footnote{\url{https://github.com/lephong/mulrel-nel}} The generation of the $P(e|m)$ index is based on the source code of~\cite{Ganea:2017:DJE} and involves the parsing of Wikipedia, the CrossWikis corpus, and YAGO. Any of these may be either removed completely, or replaced by different corpora; using the resulting $P(e|m)$ index in the package instead. %The implementation of this component enables users to replace any of these corpora and build a new index using different resources.
\\

\noindent
\emph{\textbf{ED Training.}} For the entity disambiguation method, we used the AIDA-train dataset for training and AIDA-A for validation. We use the Adam optimizer and reduce the learning rate from $1^{-3}$ to $1^{-4}$ once the F1-score of the validation set reaches $0.88$ (following~\cite{Le:2018:IEL}).
\\

\noindent
\emph{\textbf{Embeddings.}} The entity and word embeddings used for selecting candidate entities are trained on a Wikipedia 2019-07 dump using the Wikipedia2Vec package.\footnote{\url{https://wikipedia2vec.github.io/wikipedia2vec}} Following~\cite{Gerritse:2020:GEE}, we set the \emph{min-entity-count} parameter to zero and used the Wikipedia link graph during training. For the entity disambiguation model, we used GloVe embeddings~\cite{Pennington:2014:glove} as suggested in~\cite{Le:2018:IEL}.

\begin{figure}[t]
\caption{Example API input and output for entity linking.}
\label{fig:API}
\scriptsize
\begin{lstlisting}[language=json,firstnumber=1]
INPUT: 
 {"text": "Belgrade 1996-08-30 Result in an international basketball tournament on Friday: Red Star ( Yugoslavia ) beat Dinamo ( Russia) 92-90 ( halftime 47-47 )."}
\end{lstlisting}	

\begin{lstlisting}[language=json,firstnumber=1]
OUTPUT:
[
[0, 8, 'Belgrade', 'Belgrade', 0.91, 0.98, 'LOC', ], 
[80, 8, 'Red Star', 'KK_Crvena_zvezda', 0.36, 0.99, 'ORG'], 
[91, 10, 'Yugoslavia', 'Yugoslavia', 0.8, 0.99, 'LOC'], 
[109, 6, 'Dinamo', 'FC_Dinamo_Bucuresti', 0.7, 0.99, 'ORG'], 
[118, 6, 'Russia', 'Russia', 0.85, 0.99, 'LOC']
]
\end{lstlisting}
\shrink	
\end{figure}

\subsection{Usage}
REL can be used as a Python package deployed on a local machine, or as a service, via a restful API.

To use REL as a package, our GitHub repository contains step-by-step tutorials on how to perform end-to-end entity linking, and on how to (re-)train the ED model. %\todo{(s) $\Leftarrow$ as in embeddings or disambiguation model or (?); be more specific}. 
We provide scripts and instructions for deploying REL using a new Wikipedia dump; this helps REL users to keep up-to-date with emerging entities in Wikipedia, and enables researchers to deploy REL for any specific Wikipedia version that is required for a downstream task.

The API is publicly available. Given an input text, depicted in Fig.~\ref{fig:API} (Top), the API returns a list of mentions, each with (i) the start position and length of the mention, (ii) the mention itself, (iii) the linked entity, (iv) the confidence score of ED, and (vi) the confidence score and type of entity from the mention detection step (if available); see Fig.~\ref{fig:API} (Bottom).  Alternatively, a user can use the API for entity disambiguation only, by submitting an input text and a list of spans (specified with start position and length).

\section{Evaluation}
\label{sec:eval}

% GERBIL experiment links for table below:
% (EL 2014 Mick) http://gerbil.aksw.org/gerbil/experiment?id=202002140003
% (EL 2019 Mick) http://gerbil.aksw.org/gerbil/experiment?id=202002140000
% (EL 2019 AIDA + KORE50 Koen) http://gerbil.aksw.org/gerbil/experiment?id=202005250001
% (EL 2019 MSNBC Koen) http://gerbil.aksw.org/gerbil/experiment?id=202005250002
% (EL 2014 MSNBC + AIDA + KORE50 Koen) http://gerbil.aksw.org/gerbil/experiment?id=202005260000
\begin{table}[t]
\centering
\caption{EL strong matching results on the GERBIL platform.}
\label{tbl:el-gerbil}
\small
\begin{tabular}{l|@{~}l|@{~}l|@{~}l|@{~}l|@{~}l|@{~}l|@{~}l|@{~}l}
%& \rotatebox[origin=l]{90}{AIDA-A} & 
& \rotatebox[origin=l]{90}{AIDA-B} & 
\rotatebox[origin=l]{90}{MSNBC} & 
\rotatebox[origin=l]{90}{OKE-2015} & 
\rotatebox[origin=l]{90}{OKE-2016} & 
\rotatebox[origin=l]{90}{N3-Reuters-128} & 
\rotatebox[origin=l]{90}{N3-RSS-500} & 
\rotatebox[origin=l]{90}{Derczynski} & 
\rotatebox[origin=l]{90}{KORE50}
\\
\textbf{Macro F1} &&&&&&&& \\
\textbf{Micro F1} &&&&&&&&\\
\hline
DBpedia  & 52.0 & 42.4 & 42.0 & 41.4 & 21.5 & 26.7 & 33.7 & 29.4   \\
Spotlight & 57.8 & 40.6 & 44.4 & 43.1 & 24.8 & 27.2 & 32.2 & 34.9  \\
\hline
\multirow{2}*{WAT} & 70.8 & 62.6 & 53.2 & 51.8 & 45.0 & \textbf{45.3} & \textbf{44.4} & 37.3   \\
 & 73.0 & 64.5 & 56.4 & 53.9 & 49.2 & \textbf{42.3} & 38.0 & 49.6  \\
\hline
\multirow{2}*{SOTA NLP}  & \textbf{82.6} & 73.0 & 56.6 & 47.8 & 45.4 & 43.8 & 43.2 & 26.2   \\
  & 82.4 & 72.4 & 61.9 & 52.7 & \textbf{50.3} & 38.2 & 34.1 & 35.2   \\
\hline
\multirow{2}*{REL (2014)}  & 81.3 & \textbf{73.2} & 61.5 & \textbf{57.5} & \textbf{46.8} & 35.9 & 38.1 & \textbf{60.1}  \\
 & \textbf{83.3} & \textbf{74.4} & \textbf{64.8} & \textbf{58.8} & 49.7 & 34.3 & \textbf{41.2} & \textbf{61.6}  \\
\hline
\multirow{2}*{REL (2019)} & 78.6 & 71.1 & \textbf{61.8} & 57.4 & 45.7 & 36.2 & 38.0 & 50.1 \\
 & 80.5 & 72.4 & 63.1 & 58.3 & 49.9 & 35.0 & 41.1 & 50.7   \\
\hline
\end{tabular}
\miniskip
\end{table}

% GERBIL experiment links for table below:
% (ED 2014 Mick) http://gerbil.aksw.org/gerbil/experiment?id=202002130001
% (ED 2019 Mick) http://gerbil.aksw.org/gerbil/experiment?id=202002130002
\begin{table}[t]
\centering
\caption{ED results on the GERBIL platform.}
\label{tbl:ed-gerbil}
\small
\begin{tabular}{l|@{~}l|@{~}l|@{~}l|@{~}l|@{~}l|@{~}l|@{~}l|@{~}l}
%& \rotatebox[origin=l]{90}{AIDA-A} & 
& \rotatebox[origin=l]{90}{AIDA-B} & 
\rotatebox[origin=l]{90}{MSNBC} & 
\rotatebox[origin=l]{90}{OKE-2015} & 
\rotatebox[origin=l]{90}{OKE-2016} & 
\rotatebox[origin=l]{90}{N3-Reuters-128} & 
\rotatebox[origin=l]{90}{N3-RSS-500} &
\rotatebox[origin=l]{90}{Derczynski} & 
\rotatebox[origin=l]{90}{KORE50}
\\
\textbf{Macro F1} &&&&&&&\\
\textbf{Micro F1} &&&&&&&\\
\hline
DBpedia & 53.7 & 43.6 & 30.4 & 43.0 & 41.8 & 42.6 & 50.3 & 48.7  \\
Spotlight & 56.1 & 42.1 & 35.8 & 43.1 & 43.4 & 34.6 & 43.3 & 52.3 \\
\hline
\multirow{2}*{WAT} & 79.8 & 79.7 & 62.2 & 0.0 & 59.2 & 62.8 & 70.4 & 52.4 \\
 & 80.5 & 78.8 & 64.9 & 0.0 & 63.1 & 63.9 & 69.5 & 62.2  \\
\hline
\multirow{2}*{SOTA NLP} & 83.8 & 88.5 & \textbf{73.2} & \textbf{76.7} & \textbf{63.4} & \textbf{66.6} & \textbf{65.3} & 52.4 \\
 & 83.0 & 86.2 & \textbf{74.0} & \textbf{78.1} & \textbf{67.3} & \textbf{68.6} & \textbf{65.4} & 60.8  \\
\hline
 \multirow{2}*{REL (2014)} & \textbf{85.5} & \textbf{89.6} & 65.5 & 72.0 & 59.8 & 61.0 & 61.9 & \textbf{61.9}\\
  & \textbf{86.6} & \textbf{88.5} & 65.8 & 72.2 & 64.9 & 62.8 & 62.1 & \textbf{64.6}\\
 \hline
\multirow{2}*{REL (2019)} & 82.9 & 86.3 & 64.0 & 67.0 & 58.2 & 61.7 & 62.3 & 54.4 \\
 & 84.0 & 85.8 & 64.3 & 67.3 & 64.9 & 64.1 & 62.0 & 54.0  \\
\hline
\end{tabular}
\miniskip
\end{table}

We compare REL with a state-of-the-art end-to-end entity linking~\cite{Kolitsas:2018:ENE}, referred to as SOTA NLP, and two popular well-established entity linking systems: (i) DBpedia-spotlight~\cite{Mendes:2011:DSS} and (ii) WAT~\cite{Piccinno:2014:WAT}, the updated version of \textsc{TagMe}~\cite{Ferragina:2010:TOA}.
We report the results for two versions of our system. The first one, denoted as \emph{REL (2014)}, is based on the original implementation of~\cite{Le:2018:IEL} for ED.  It uses Wikipedia 2014 as the reference knowledge base and employs entity embeddings provided by~\cite{Ganea:2017:DJE} for candidate selection. The second version of our system, denoted as \emph{REL (2019)}, is based on Wikipedia 2019-07 and uses Wikipedia2Vec embeddings; cf. Section~\ref{sec:Imp}.

We use the GERBIL platform~\cite{Roder:2018:Gerbil} for evaluation, and report on micro and macro InKB F1 scores for both EL and ED. %We note that AIDA-A dataset is used as validation set for SOTA NLP and REL. Therefore, AIDA-A results cannot be used for a fair comparison between systems.
Table~\ref{tbl:el-gerbil} shows the strong matching results for EL, where strong refers to the requirement of  exactly predicting the gold mention boundaries. We first note that REL outperforms the well-established entity linking toolkits (DBpedia Spotlight and WAT) by a large margin. Comparing with SOTA NLP, we observe that REL (2019) outperforms (or performs on par with) SOTA NLP on half of the datasets. 
The ED results in Table~\ref{tbl:ed-gerbil} also show consistent and significant improvements of REL over the two well-established toolkits. SOTA NLP, however, obtains better results than REL for all, except three datasets.
For both EL and ED results, we observe that REL (2014) achieves better results compared to REL (2019). This can be attributed to the different embeddings used for candidate selection: the recall of candidate entities chosen by their similarity to the context of the mentions is lower in REL (2019) when compared to REL (2014).

For a reference comparison, we also report the results of the ED method (referred to as MulRel-NEL) as reported in~\cite{Le:2018:IEL}; see Table~\ref{tbl:el_local}. The micro F1 score reported in this table is computed locally and by matching ED results against the original datasets. The results show that REL (2014) and MulRel-NEL scores are almost identical, which attests to the repeatability of~\cite{Le:2018:IEL}. Again, we observe a decrease in performance when comparing REL (2019) to REL (2014), just like in Table~\ref{tbl:ed-gerbil}.

Finally, we report on the runtime efficiency of REL in Table~\ref{tbl:efficiency}.  Specifically, we measure efficiency on a random sample of 50 documents (with a minimum length of 200 words) taken from AIDA-B.  
The experiments were run on a laptop with Intel i7 CPU (2.80GHz), 16GB RAM, and an NVIDIA Geforce GTX 1050 (4GB) GPU. The results show that detecting the mentions takes considerably more time than ED, and is done more efficiently using GPU. The ED time, however, is less affected by the GPU usage. This indicates that the overall efficiency of REL can be improved by replacing MD with a more efficient NER approach.
% in two settings: \emph{per-doc EL}, which is the average annotation time when each document is annotated separately, and \emph{batch EL}, where is total annotation time when all documents are annotated in one batch.

\begin{table}[t]
\centering
\caption{Local ED results as reported in \cite{Le:2018:IEL}}
\label{tbl:el_local}
\small
% \resizebox{\textwidth}{!}{%
\begin{tabular}{l|l|l|l|l|l|l}
%\hline
& \rotatebox[origin=l]{90}{AIDA-B} &
\rotatebox[origin=l]{90}{ACE2004} &
\rotatebox[origin=l]{90}{Aquaint} &
\rotatebox[origin=l]{90}{CLUEWEB} &
\rotatebox[origin=l]{90}{MSNBC} &
\rotatebox[origin=l]{90}{Wikipedia} \\
\textbf{Micro F1} &&&&&& \\
\hline
MulRel-NEL~\cite{Le:2018:IEL} & 93.1 & 89.9 & 88.3 & 77.5 & 93.9 & 78.0
\\
\hline
REL (2014) & 92.8 & 89.7 & 87.4 & 77.6 & 93.5 & 78.7 \\
\hline
REL (2019) & 89.4 & 85.3 & 84.1 & 71.9 & 90.7 & 73.1 \\
\hline
\end{tabular}
\miniskip
\end{table}

%\begin{table*}[t]
%\centering
%\caption{ED GsInKb results}
%\label{tab:ed_gerbil_gsinb}
%% \resizebox{\textwidth}{!}{%
%\begin{tabular}{llllllllll}
%\hline
%& \rotatebox[origin=c]{90}{AIDA A} & \rotatebox[origin=c]{90}{AIDA B} & \rotatebox[origin=c]{90}{MSNBC} & \rotatebox[origin=c]{90}{OKE-2015} & \rotatebox[origin=c]{90}{OKE-2016} & \rotatebox[origin=c]{90}{N3-Reuters-128} & \rotatebox[origin=c]{90}{N3-RSS-500} & \rotatebox[origin=c]{90}{Derczynski} & \rotatebox[origin=c]{90}{KORE50} \\ 
%F1@MA  \\
%F1@MI \\
%\hline
%\multirow{2}*{REL '14} & 90.0 & 92.7 & 93.4 & 72.5 & 80.0 & 66.3 & 79.1 & 78.8 & 62.1 \\
% & 91.3 & 92.3 & 91.5 & 71.2 & 77.8 & 69.0 & 72.1 & 73.6 & 64.8\\
% \hline
%\multirow{2}*{REL '19} & 86.9 & 89.9 & 90.2 & 71.1 & 73.9 & 65.0 & 80.5 & 78.9 & 52.6 \\
% & 88.5 & 89.7 & 88.9 & 69.7 & 72.5 & 69.2 & 74.4 & 73.6 & 54.2  \\
%\hline
%\end{tabular}%
%\end{table*}

\begin{table}[t]
\centering
\caption{Efficiency of REL (in seconds) for 50 documents from AIDA-B with $>$ 200 words, which is 323 ($\pm$ 105) words and 42 ($\pm$ 19) mentions per document.}
\label{tbl:efficiency}
\small
% \resizebox{\textwidth}{!}{%
\begin{tabular}{l|ll}
\hline
%& \multicolumn{2}{c|}{\textbf{With GPU}} & \multicolumn{2}{c}{\textbf{Without GPU}}
& \textbf{Time MD} & \textbf{Time ED} \\
\hline
With GPU & 0.44$\pm$0.22 & 0.24$\pm$0.08   \\
Without GPU & 2.41$\pm$1.24 & 0.18$\pm$0.09 \\
\hline
\end{tabular}
\miniskip
\end{table}

%\begin{table}[t]
%\centering
%\caption{Statistics efficiency per document originating from AIDA-A and AIDA-B using Wiki 2019 (with GPU).}
%\label{tab:efficiency_e2e_gpu}
%% \resizebox{\textwidth}{!}{%
%\begin{tabular}{llllllllll}
%\hline
%& \#words & Mean \#mentions & Time MD (s) & Time ED (s) \\ 
%\hline
%Mean & 192 & 26 & 0.2844 & 0.0524 \\
%St. dev. & 156 & 25 & 0.2252 & 0.0517 \\
%\hline
%\end{tabular}%
%% }
%\end{table}
%
%\begin{table}[t]
%\centering
%\caption{Statistics efficiency per document originating from AIDA-A and AIDA-B using Wiki 2019 (without CPU).}
%\label{tab:efficiency_e2e_cpu}
%% \resizebox{\textwidth}{!}{%
%\begin{tabular}{llllllllll}
%\hline
%& n words & Mean n mentions & Time MD (s) & Time ED (s) \\ 
%\hline
%Mean & 192 & 26 & 1.5800 & 0.0659 \\
%St. dev. & 156 & 25 & 1.1826 & 0.1008 \\
%\hline
%\end{tabular}%
%% }
%\end{table}

\section{Conclusion}
We have introduced the Radboud Entity Linker (REL), an open source toolkit for entity linking. REL builds on state-of-the-art neural components from natural language processing research, and is provided as a Python package and as a web API. Currently, REL is optimized for annotating documents and short texts. In the future,  we plan to train REL on a large corpus of annotated queries and make it available for the task of entity linking in queries as well.

%\paragraph{\textbf{Acknowledgements}} 

\bibliographystyle{ACM-Reference-Format}
\bibliography{ELdemo}


\begin{thebibliography}{00}

%%% the Web address.

\ifx \showCODEN    \undefined \def \showCODEN     #1{\unskip}     \fi
\ifx \showDOI      \undefined \def \showDOI       #1{{\tt DOI:}\penalty0{#1}\ }
  \fi
\ifx \showISBNx    \undefined \def \showISBNx     #1{\unskip}     \fi
\ifx \showISBNxiii \undefined \def \showISBNxiii  #1{\unskip}     \fi
\ifx \showISSN     \undefined \def \showISSN      #1{\unskip}     \fi
\ifx \showLCCN     \undefined \def \showLCCN      #1{\unskip}     \fi
\ifx \shownote     \undefined \def \shownote      #1{#1}          \fi
\ifx \showarticletitle \undefined \def \showarticletitle #1{#1}   \fi
\ifx \showURL      \undefined \def \showURL       #1{#1}          \fi
% invisible to TeX
\providecommand\bibfield[2]{#2}
\providecommand\bibinfo[2]{#2}
\providecommand\natexlab[1]{#1}
\providecommand\showeprint[2][]{arXiv:#2}

\bibitem[\protect\citeauthoryear{Akbik, Blythe, and Vollgraf}{Akbik
  et~al\mbox{.}}{2018}]%
        {Akbik:2018:flair}
\bibfield{author}{\bibinfo{person}{Alan Akbik}, \bibinfo{person}{Duncan
  Blythe}, {and} \bibinfo{person}{Roland Vollgraf}.}
  \bibinfo{year}{2018}\natexlab{}.
\newblock \showarticletitle{Contextual String Embeddings for Sequence
  Labeling}. In \bibinfo{booktitle}{{\em Proc. of COLING '18}}.
  \bibinfo{pages}{1638--1649}.
\newblock


\bibitem[\protect\citeauthoryear{Balog}{Balog}{2018}]%
        {Balog:2018:EOS}
\bibfield{author}{\bibinfo{person}{Krisztian Balog}.}
  \bibinfo{year}{2018}\natexlab{}.
\newblock \bibinfo{booktitle}{{\em Entity-Oriented Search}}.
  \bibinfo{series}{The Information Retrieval Series},
  Vol.~\bibinfo{volume}{39}.
\newblock \bibinfo{publisher}{Springer}.
\newblock


\bibitem[\protect\citeauthoryear{Balog, Ramampiaro, Takhirov, and
  N{\o}rv{\aa}g}{Balog et~al\mbox{.}}{2013}]%
        {Balog:2013:MCA}
\bibfield{author}{\bibinfo{person}{Krisztian Balog}, \bibinfo{person}{Heri
  Ramampiaro}, \bibinfo{person}{Naimdjon Takhirov}, {and}
  \bibinfo{person}{Kjetil N{\o}rv{\aa}g}.} \bibinfo{year}{2013}\natexlab{}.
\newblock \showarticletitle{Multi-step Classification Approaches to Cumulative
  Citation Recommendation}. In \bibinfo{booktitle}{{\em Proc. of OAIR '13}}.
  \bibinfo{pages}{121--128}.
\newblock


\bibitem[\protect\citeauthoryear{Bunescu and Pa\c{s}ca}{Bunescu and
  Pa\c{s}ca}{2006}]%
        {Bunescu:2006:UEK}
\bibfield{author}{\bibinfo{person}{Razvan Bunescu} {and}
  \bibinfo{person}{Marius Pa\c{s}ca}.} \bibinfo{year}{2006}\natexlab{}.
\newblock \showarticletitle{Using Encyclopedic Knowledge for Named Entity
  Disambiguation}. In \bibinfo{booktitle}{{\em Proc. of EACL '06}}.
  \bibinfo{pages}{9--16}.
\newblock


\bibitem[\protect\citeauthoryear{Cornolti, Ferragina, Ciaramita, R\"{u}d, and
  Sch\"{u}tze}{Cornolti et~al\mbox{.}}{2018}]%
        {Cornolti:2018:APA}
\bibfield{author}{\bibinfo{person}{Marco Cornolti}, \bibinfo{person}{Paolo
  Ferragina}, \bibinfo{person}{Massimiliano Ciaramita}, \bibinfo{person}{Stefan
  R\"{u}d}, {and} \bibinfo{person}{Hinrich Sch\"{u}tze}.}
  \bibinfo{year}{2018}\natexlab{}.
\newblock \showarticletitle{SMAPH: A Piggyback Approach for Entity-Linking in
  Web Queries}.
\newblock \bibinfo{journal}{{\em ACM Trans. Inf. Syst.\/}}
  \bibinfo{volume}{37}, \bibinfo{number}{1} (\bibinfo{year}{2018}).
\newblock


\bibitem[\protect\citeauthoryear{Cucerzan}{Cucerzan}{2007}]%
        {Cucerzan:2007:LSN}
\bibfield{author}{\bibinfo{person}{Silviu Cucerzan}.}
  \bibinfo{year}{2007}\natexlab{}.
\newblock \showarticletitle{Large-Scale Named Entity Disambiguation Based on
  {W}ikipedia Data}. In \bibinfo{booktitle}{{\em Proc. of EMNLP-CoNLL '07}}.
  \bibinfo{pages}{708--716}.
\newblock


\bibitem[\protect\citeauthoryear{Ferragina and Scaiella}{Ferragina and
  Scaiella}{2010}]%
        {Ferragina:2010:TOA}
\bibfield{author}{\bibinfo{person}{Paolo Ferragina} {and} \bibinfo{person}{Ugo
  Scaiella}.} \bibinfo{year}{2010}\natexlab{}.
\newblock \showarticletitle{{TAGME:} {O}n-the-fly Annotation of Short Text
  Fragments (by {W}ikipedia Entities)}. In \bibinfo{booktitle}{{\em Proc. of
  CIKM '10}}. \bibinfo{pages}{1625--1628}.
\newblock


\bibitem[\protect\citeauthoryear{Ganea and Hofmann}{Ganea and Hofmann}{2017}]%
        {Ganea:2017:DJE}
\bibfield{author}{\bibinfo{person}{Octavian-Eugen Ganea} {and}
  \bibinfo{person}{Thomas Hofmann}.} \bibinfo{year}{2017}\natexlab{}.
\newblock \showarticletitle{Deep Joint Entity Disambiguation with Local Neural
  Attention}. In \bibinfo{booktitle}{{\em Proc. of EMNLP '17}}.
  \bibinfo{pages}{2619--2629}.
\newblock


\bibitem[\protect\citeauthoryear{Gerritse, Hasibi, and de~Vries}{Gerritse
  et~al\mbox{.}}{}]%
        {Gerritse:2020:GEE}
\bibfield{author}{\bibinfo{person}{Emma Gerritse}, \bibinfo{person}{Faegheh
  Hasibi}, {and} \bibinfo{person}{Arjen~P. de Vries}.}
\newblock \showarticletitle{Graph-Embedding Empowered Entity Retrieval}. In
  \bibinfo{booktitle}{{\em Proc. of ECIR '20}}. \bibinfo{pages}{97--110}.
\newblock


\bibitem[\protect\citeauthoryear{Hasibi, Balog, and Bratsberg}{Hasibi
  et~al\mbox{.}}{2015}]%
        {Hasibi:2015:ELQ}
\bibfield{author}{\bibinfo{person}{Faegheh Hasibi}, \bibinfo{person}{Krisztian
  Balog}, {and} \bibinfo{person}{Svein~Erik Bratsberg}.}
  \bibinfo{year}{2015}\natexlab{}.
\newblock \showarticletitle{Entity Linking in Queries: Tasks and Evaluation}.
  In \bibinfo{booktitle}{{\em Proc. of ICTIR '15}}. \bibinfo{pages}{171--180}.
\newblock


\bibitem[\protect\citeauthoryear{Hasibi, Balog, and Bratsberg}{Hasibi
  et~al\mbox{.}}{2016}]%
        {Hasibi:2016:EEL}
\bibfield{author}{\bibinfo{person}{Faegheh Hasibi}, \bibinfo{person}{Krisztian
  Balog}, {and} \bibinfo{person}{Svein~Erik Bratsberg}.}
  \bibinfo{year}{2016}\natexlab{}.
\newblock \showarticletitle{Exploiting Entity Linking in Queries for Entity
  Retrieval}. In \bibinfo{booktitle}{{\em Proc. of ICTIR '16}}.
  \bibinfo{pages}{209--218}.
\newblock


\bibitem[\protect\citeauthoryear{Hasibi, Balog, Garigliotti, and Zhang}{Hasibi
  et~al\mbox{.}}{2017}]%
        {Hasibi:2017:NTE}
\bibfield{author}{\bibinfo{person}{Faegheh Hasibi}, \bibinfo{person}{Krisztian
  Balog}, \bibinfo{person}{Dar\'{\i}o Garigliotti}, {and} \bibinfo{person}{Shuo
  Zhang}.} \bibinfo{year}{2017}\natexlab{}.
\newblock \showarticletitle{Nordlys: A Toolkit for Entity-Oriented and Semantic
  Search}. In \bibinfo{booktitle}{{\em Proc. of SIGIR '17}}.
  \bibinfo{pages}{1289--1292}.
\newblock


\bibitem[\protect\citeauthoryear{Hoffart, Yosef, Bordino, F{\"{u}}rstenau,
  Pinkal, Spaniol, Taneva, Thater, and Weikum}{Hoffart et~al\mbox{.}}{2011}]%
        {Hoffart:2011:RDN}
\bibfield{author}{\bibinfo{person}{Johannes Hoffart},
  \bibinfo{person}{Mohamed~Amir Yosef}, \bibinfo{person}{Ilaria Bordino},
  \bibinfo{person}{Hagen F{\"{u}}rstenau}, \bibinfo{person}{Manfred Pinkal},
  \bibinfo{person}{Marc Spaniol}, \bibinfo{person}{Bilyana Taneva},
  \bibinfo{person}{Stefan Thater}, {and} \bibinfo{person}{Gerhard Weikum}.}
  \bibinfo{year}{2011}\natexlab{}.
\newblock \showarticletitle{Robust Disambiguation of Named Entities in Text}.
  In \bibinfo{booktitle}{{\em Proc. of EMNLP '11}}. \bibinfo{pages}{782--792}.
\newblock


\bibitem[\protect\citeauthoryear{Kolitsas, Ganea, and Hofmann}{Kolitsas
  et~al\mbox{.}}{2018}]%
        {Kolitsas:2018:ENE}
\bibfield{author}{\bibinfo{person}{Nikolaos Kolitsas},
  \bibinfo{person}{Octavian-Eugen Ganea}, {and} \bibinfo{person}{Thomas
  Hofmann}.} \bibinfo{year}{2018}\natexlab{}.
\newblock \showarticletitle{End-to-End Neural Entity Linking}. In
  \bibinfo{booktitle}{{\em Proc. of CoNLL '18}}. \bibinfo{pages}{519--529}.
\newblock


\bibitem[\protect\citeauthoryear{Le and Titov}{Le and Titov}{2018}]%
        {Le:2018:IEL}
\bibfield{author}{\bibinfo{person}{Phong Le} {and} \bibinfo{person}{Ivan
  Titov}.} \bibinfo{year}{2018}\natexlab{}.
\newblock \showarticletitle{Improving Entity Linking by Modeling Latent
  Relations between Mentions}. In \bibinfo{booktitle}{{\em Proc. of ACL '18}}.
  \bibinfo{pages}{1595--1604}.
\newblock


\bibitem[\protect\citeauthoryear{Mendes, Jakob, Garc{\'{i}}a-Silva, and
  Bizer}{Mendes et~al\mbox{.}}{2011}]%
        {Mendes:2011:DSS}
\bibfield{author}{\bibinfo{person}{Pablo~N Mendes}, \bibinfo{person}{Max
  Jakob}, \bibinfo{person}{Andr{\'{e}}s Garc{\'{i}}a-Silva}, {and}
  \bibinfo{person}{Christian Bizer}.} \bibinfo{year}{2011}\natexlab{}.
\newblock \showarticletitle{DBpedia {S}potlight: Shedding Light on the Web of
  Documents}. In \bibinfo{booktitle}{{\em Proc. of I-Semantics '11}}.
  \bibinfo{pages}{1--8}.
\newblock


\bibitem[\protect\citeauthoryear{Mihalcea and Csomai}{Mihalcea and
  Csomai}{2007}]%
        {Mihalcea:2007:WLD}
\bibfield{author}{\bibinfo{person}{Rada Mihalcea} {and} \bibinfo{person}{Andras
  Csomai}.} \bibinfo{year}{2007}\natexlab{}.
\newblock \showarticletitle{Wikify! - {L}inking Documents to Encyclopedic
  Knowledge}. In \bibinfo{booktitle}{{\em Proc. of CIKM '07}}.
  \bibinfo{pages}{233--242}.
\newblock


\bibitem[\protect\citeauthoryear{Milne and Witten}{Milne and Witten}{2008}]%
        {Milne:2008:LLW}
\bibfield{author}{\bibinfo{person}{David Milne} {and} \bibinfo{person}{Ian~H
  Witten}.} \bibinfo{year}{2008}\natexlab{}.
\newblock \showarticletitle{Learning to Link with {W}ikipedia}. In
  \bibinfo{booktitle}{{\em Proc. of CIKM '08}}. \bibinfo{pages}{509--518}.
\newblock


\bibitem[\protect\citeauthoryear{Pappu, Blanco, Mehdad, Stent, and
  Thadani}{Pappu et~al\mbox{.}}{2017}]%
        {Pappu:2017:LME}
\bibfield{author}{\bibinfo{person}{Aasish Pappu}, \bibinfo{person}{Roi Blanco},
  \bibinfo{person}{Yashar Mehdad}, \bibinfo{person}{Amanda Stent}, {and}
  \bibinfo{person}{Kapil Thadani}.} \bibinfo{year}{2017}\natexlab{}.
\newblock \showarticletitle{Lightweight Multilingual Entity Extraction and
  Linking}. In \bibinfo{booktitle}{{\em Proc. of WSDM '17}}.
  \bibinfo{pages}{365--374}.
\newblock


\bibitem[\protect\citeauthoryear{Pennington, Socher, and Manning}{Pennington
  et~al\mbox{.}}{2014}]%
        {Pennington:2014:glove}
\bibfield{author}{\bibinfo{person}{Jeffrey Pennington},
  \bibinfo{person}{Richard Socher}, {and} \bibinfo{person}{Christopher
  Manning}.} \bibinfo{year}{2014}\natexlab{}.
\newblock \showarticletitle{{G}love: Global Vectors for Word Representation}.
  In \bibinfo{booktitle}{{\em Proc. of EMNLP '14}}.
  \bibinfo{pages}{1532--1543}.
\newblock


\bibitem[\protect\citeauthoryear{Piccinno, Ferragina, and Informatica}{Piccinno
  et~al\mbox{.}}{2014}]%
        {Piccinno:2014:WAT}
\bibfield{author}{\bibinfo{person}{Francesco Piccinno}, \bibinfo{person}{Paolo
  Ferragina}, {and} \bibinfo{person}{Dipartimento Informatica}.}
  \bibinfo{year}{2014}\natexlab{}.
\newblock \showarticletitle{{From TagME to WAT} : a new Entity Annotator
  Categories and Subject Descriptors}.
\newblock  (\bibinfo{year}{2014}), \bibinfo{pages}{55--62}.
\newblock


\bibitem[\protect\citeauthoryear{Platt}{Platt}{2000}]%
        {Platt:2000:PSVM}
\bibfield{author}{\bibinfo{person}{John Platt}.}
  \bibinfo{year}{2000}\natexlab{}.
\newblock \showarticletitle{Probabilities for SV Machines}. In
  \bibinfo{booktitle}{{\em Advances in Large-Margin Classifiers}}.
  \bibinfo{pages}{61--73}.
\newblock


\bibitem[\protect\citeauthoryear{Reinanda, Meij, and de~Rijke}{Reinanda
  et~al\mbox{.}}{2015}]%
        {Reinanda:2015:MRR}
\bibfield{author}{\bibinfo{person}{Ridho Reinanda}, \bibinfo{person}{Edgar
  Meij}, {and} \bibinfo{person}{Maarten de Rijke}.}
  \bibinfo{year}{2015}\natexlab{}.
\newblock \showarticletitle{Mining, Ranking and Recommending Entity Aspects}.
  In \bibinfo{booktitle}{{\em Proc. of SIGIR '15}}. \bibinfo{pages}{263--272}.
\newblock


\bibitem[\protect\citeauthoryear{R{\"o}der, Usbeck, and Ngonga~Ngomo}{R{\"o}der
  et~al\mbox{.}}{2018}]%
        {Roder:2018:Gerbil}
\bibfield{author}{\bibinfo{person}{Michael R{\"o}der}, \bibinfo{person}{Ricardo
  Usbeck}, {and} \bibinfo{person}{Axel-Cyrille Ngonga~Ngomo}.}
  \bibinfo{year}{2018}\natexlab{}.
\newblock \showarticletitle{GERBIL--Benchmarking Named Entity Recognition and
  Linking Consistently}.
\newblock \bibinfo{journal}{{\em Semantic Web\/}} \bibinfo{volume}{9},
  \bibinfo{number}{5} (\bibinfo{year}{2018}), \bibinfo{pages}{605--625}.
\newblock


\bibitem[\protect\citeauthoryear{Spitkovsky and Chang}{Spitkovsky and
  Chang}{2012}]%
        {Spitkovsky:2012:CDE}
\bibfield{author}{\bibinfo{person}{Valentin~I. Spitkovsky} {and}
  \bibinfo{person}{Angel~X. Chang}.} \bibinfo{year}{2012}\natexlab{}.
\newblock \showarticletitle{A Cross-Lingual Dictionary for {E}nglish
  {W}ikipedia Concepts}. In \bibinfo{booktitle}{{\em Proc. of LREC'12}}.
  \bibinfo{pages}{3168--3175}.
\newblock


\bibitem[\protect\citeauthoryear{Xiong, Callan, and Liu}{Xiong
  et~al\mbox{.}}{2017}]%
        {Xiong:2017:WDR}
\bibfield{author}{\bibinfo{person}{Chenyan Xiong}, \bibinfo{person}{Jamie
  Callan}, {and} \bibinfo{person}{Tie-Yan Liu}.}
  \bibinfo{year}{2017}\natexlab{}.
\newblock \showarticletitle{Word-Entity Duet Representations for Document
  Ranking}. In \bibinfo{booktitle}{{\em Proc. of SIGIR '17}}.
  \bibinfo{pages}{763--772}.
\newblock


\bibitem[\protect\citeauthoryear{Yamada, Asai, Shindo, Takeda, and
  Takefuji}{Yamada et~al\mbox{.}}{2018}]%
        {Yamada:2018:WOT}
\bibfield{author}{\bibinfo{person}{Ikuya Yamada}, \bibinfo{person}{Akari Asai},
  \bibinfo{person}{Hiroyuki Shindo}, \bibinfo{person}{Hideaki Takeda}, {and}
  \bibinfo{person}{Yoshiyasu Takefuji}.} \bibinfo{year}{2018}\natexlab{}.
\newblock \showarticletitle{Wikipedia2Vec: An Optimized Tool for Learning
  Embeddings of Words and Entities from Wikipedia}.
\newblock \bibinfo{journal}{{\em arXiv preprint 1812.06280\/}}
  (\bibinfo{year}{2018}).
\newblock


\bibitem[\protect\citeauthoryear{Yamada, Shindo, Takeda, and Takefuji}{Yamada
  et~al\mbox{.}}{2016}]%
        {Yamada:2016:JLE}
\bibfield{author}{\bibinfo{person}{Ikuya Yamada}, \bibinfo{person}{Hiroyuki
  Shindo}, \bibinfo{person}{Hideaki Takeda}, {and} \bibinfo{person}{Yoshiyasu
  Takefuji}.} \bibinfo{year}{2016}\natexlab{}.
\newblock \showarticletitle{Joint learning of the embedding of words and
  entities for named entity disambiguation}. In \bibinfo{booktitle}{{\em Proc
  of CoNLL '16}}. \bibinfo{pages}{250--259}.
\newblock


\end{thebibliography}

\end{document}